# Context, input and process as critical elements for successful Emergency Remote Learning

Luciana Oliveira[1[0000-0003-2419-4332]], Anabela Mesquita[2[0000-0001-8564-4582]], Arminda Sequeira[1[0000-0003-1457-6070]], Adriana Oliveira[1[0000-0003-0081-2335]] and Paulino Silva[1[0000-0003-1443-4961]]

[1] CEOS.PP ISCAP Polytechnic of Porto, Rua Jaime Lopes Amorim, Matosinhos, Portugal
[2] Polytechnic of Porto & Algoritmi RC, Rua Jaime Lopes Amorim, Matosinhos, Portugal
lgo@iscap.ipp.pt

**Abstract.** In Spring 2020, the world moved from traditional classes to what was coined as ERL (Emergency Remote Teaching / Learning / Instruction), posing real challenges to all actors involved, requiring an immediate, unprecedented, and unplanned devising of mitigation strategies. The impacts of this transition cannot, however, be studied only at the educational level, as it consists of a broader social shift with multidomain repercussions. In this paper, we use the CIPP model (Context, Input, Process and Product evaluations) to further investigate interrelations among the context, input and process elements of ERL during the first wave of COVID-19, as the second wave presses towards reconfining. A correlation analysis of 46 variables, based students' responses (N=360) to a closed-ended questionnaire shows the crucial importance of motivation and engagement in online classes, as learning enablers or constrainers. These also shape the students' perception of the role that online classes play in helping them to stay more positive during ERL.

## 1 Introduction

In March 2020, the world moved from traditional classes to what was coined as ERL (Emergency Remote Teaching / Learning / Instruction). The urgent and unplanned transition to fully remote teaching posed real challenges to all actors involved in the process, requiring immediate and unprecedented use of technologies, in particular in education. The impacts of this transition cannot, however, be studied only at the educational level, as it consists of a broader social shift with multilevel repercussions. In this paper, we use the CIPP model – Context, Input, Process and Product evaluations [1] to reflect on the circumstances and impact of the Emergency Remote Learning. The context assesses needs, problems, assets, and opportunities, as well as relevant contextual conditions and dynamics (e.g., institutional, social, financial, and governmental aspects). The input evaluation focus on how it should be done and assesses competing strategies and the work plans and budgets of the selected approach (e.g, technology infrastructure, software, faculty support, faculty professional develop-ment, learning resources such as access to libraries). The process evaluation considers if the program is being done and consists of monitoring, documenting, assessing, and reporting on the implementation of plans (e.g., how processes can be adapted to this new reality, quality of teaching and learning). Finally, the product evaluation sees if the program succeeded and it identifies and assesses costs and outcomes – intended and unintended short term and long term (e.g., course completion rates, aggregated grade analyses, feedback).

According to Hodges [2], the "Evaluation of ERT should be more focused on the context, input, and process elements than product (learning)". The author explains that they "are not advocating for no evaluation of whether or not learning occurred, or to what extent it occurred, but simply stressing that the urgency of ERT and all that will take to make it happen in a short time frame will be the most critical elements to evaluate during this crisis". As such, in previous research, we focused on reporting on the overall evaluation according to six proposed dimensions. In this paper, we focus on the CIPP model's context, input and process elements concerning in Northern Portugal. As of October 2020, with the rise of the second wave of COVID-19, and the pressure to reconfine, we further investigate which are the most impactful variables of these CIPP Model elements, and how they interrelate, in order to obtain in-depth knowledge to foster current strategies being devised by teachers and Higher Education Institutions.

## 2 Background: the CIPP model

In this paper we focus on the CIPP model's context, input and process elements, leaving the process element out, according to what is recommended by Hodges [2].

### 2.1 Context

The first lockdown and the consequent adoption of ERL forced nearly 2 billion students worldwide to move online. One of the contextual challenges felt is related to this years' university graduates [3] as these experienced major interruptions in teaching and assessment in the final stage of education. This may imply, for some, postponing graduation, and job market entry. This situation is aggravated by the global recession caused by COVID-19, with increased unemployment rates and salary reduction, which tends to get worse as the second wave expands.

Family issues should also be considered, pertaining to two different scopes: the structure and stability of family income and, the conditions of the physical space that the family shares, which may be very limited, raising questions of privacy but also of equipment (desktops, laptops, or iPads) and internet access sharing. If the family is experiencing income shrinking, the continuation of studies may be threatened. The instability of family income is also linked to students' anxiety, which has been considered one of the most evident psychological problems [3, 4].

These psychological issues had their origin in the feeling of uncertainty about what was going to happen, namely, on the studies [5], future employment [6], and previous psychological health of students [7, 8]. It could also have been caused by the gradually increasing distance between people resulting from the quarantine that affected all students and in particular, those staying far from home as they were not only worried about their health, safety, and education but they also had concerns for the wellbeing of their families. Furthermore, anxiety disorders are more likely to occur and worsen in the absence of interpersonal communication [9, 10], which was the case. The significant shortage of masks and disinfectants during the first wave, the overwhelming and sensational news headlines, and erroneous news reports also contributed to this effect [11].

Studies also show that the anxiety regarding the epidemic was associated with the place of residence of the students, source of parental income, whether living with parents and whether a relative or an acquaintance was infected with COVID-19 [4]. Living with parents is a favourable factor against feeling anxious. Moreover, social support not only reduces the psychological pressure during the epidemics but also changes the attitude regarding social support and help-seeking methods. This result suggests that effective and robust social support is necessary during public health emergencies [12].

In some countries, for some individuals, the psychological impact of the pandemic situation was suggested to have been more significant than the physical health danger posed by the diseases themselves [13] affecting the mental health of college students.

In addition to the psychological factors described above, it is necessary to consider that the economic crisis and the sanitary measures to combat the epidemic will produce a significant personal income shrinkage [14]. This might result on an increase of the number of working students, as it is often college students who report the highest levels of financial strain, defined by their perceptions of economic stress or lack of financial support from their families, that feel most compelled to work during their undergraduate studies [15, 16]. Additionally, as distance education does not require the student to move to the city/region where the university is located, this means that those who previously could not afford or did not want to relocate can now apply for admission in a university in another region [14]. Finally, it is known that students' financial situation has a high impact on a student's dropout decision [17].

Another aspect to consider relates to lecturers as they are an important part of the educational ecosystem, designed to support learners with formal, informal, and social resources. Any efficient online education requires an investment in an ecosystem of learners' support, which takes time to identify and build. Of course, it is possible to use a simple online content delivery which can be quick and inexpensive, although it can be confusing and not robust. This means it is essential to build a whole ecosystem integrating all sorts of instruction in which the planning of the design process is key [2]. Issues of online design must be taken into consideration as well. In a period of crisis and crisis response, there is an increased risk of diminishing the quality of the courses delivered. If there was already a stigma concerning the online instruction, with people thinking it has lower quality, this prejudice might be exacerbated with this quick move online, without the proper preparation of all the key actors involved. Usually, a full-course development project can take months when done correctly. The need to "just get it online" is in direct contradiction to the time and effort generally dedicated to developing quality education.

### 2.2 Input

The input concerns the resources available to implement the ERL during the COVID-19 crisis. One of the categories of these is the technology as it is central for both teachers and students although less accessible for students, as some of them do not have computers, and others may have to share the computer they have, as parents and siblings are also working remotely during the crisis. Furthermore, the lack of internet access is exacerbating the digital divide [2].

### 2.3 Process

One of the challenges during the first lockdown in the Spring of 2020 was the loss of face-to-face contact and direct interactions with both colleagues and teachers and, some studies suggest that students experienced severe limitations in subjects that benefited from physical interaction with the materials, and tended to lose the ”pacing mechanism” of scheduled lectures, thus having a higher chance of dropping out than those in traditional settings [18, 19].

Another challenge is related to the interaction. Knowledge is socially created [20], and this means that it hatches from interactions between students, students and the content and students and the teacher to increases the learning outcomes. Social constructionism recognizes learning as both a social and a cognitive process, not merely a transmission of information [2]. Chat rooms or real-time tutorials were used to maximize interactions.

One lesson learned with the first lockdown was the importance of empathy in ERL, as the “ability to understand and share the feeling of another” [21]. According to the results of the survey, which compared the perspectives of teachers and students, for faculty, the remote teaching and learning experience was much more positive to students. During and after the lockdown, they say they learn better in traditional classes and that ERL was less effective [22]. They claimed to have missed discussions, engagement, and interaction. Another important issue relies on providing feedback to students. As stated, “Anything over a week after the students submit their work may be too long” [21]. Equity and inclusion are also relevant for the success of online instruction, as stated by Jebsen “The intent is to improve access to learning more equally while treating everyone equitably” [23].

Moreover, students may also face other challenges, such as lack of a structured environment that prevents them from remaining motivated or focused [24]. Also, as students are at home, they can suffer from negative emotions and poor mental health (anxiety and depression is a result of missing friends and isolation). Physical problems of being online all the time (such as strained eyes and migraines) should also be added to the list. For all these reasons, it is important that lecturers are able to create a safe and comfortable educational online environment and develop empathy with all their students.

## 3 Methodology

The adopted methodology consists of exploratory survey-based research, quantitative in nature. The authors proposed an evaluation instrument [22], a close-ended questionnaire, comprised of 67 variables organized in six dimensions of issues that impact ERL, as follows: (A) Educational and organizational issues – 33 items, (B) Technological and working conditions – 5 items, (C) Social issues – 11 items, (D) Family-related issues – 4 items, (E) Psychological issues – 6 items, (F) Financial issues – 8 items, framed in the context, input and process elements of the CIPP model. Each item on the questionnaire was presented in a labelled 4-point Likert scale anchored at 1 = never, 2 = rarely, 3 = frequently, and 4 = always.

The main goal of this stage of analysis was to identify the most impactful variables, among the initial set of 67, by expressing their interdependencies and adding to previous knowledge about students' perceptions regarding ERL.

The survey was disseminated among students enrolled in HEI in Northern Portugal, through social media channels, namely in institutional public pages and open groups. Survey data were downloaded and transferred into IBM SPSS Statistics 26.0, and the correlation analysis was transposed into a network developed in Gephi 0.9.2. The sample consists of 360 valid responses from students, from which we considered only the 341 students that indicated they were involved in ERL. The great majority of the participants are female (74.72%), aged 18 to 22 years (83.06%).

## 4 Results

A series of Spearman rank-order correlations were conducted to determine the relationships among the 67 categorical nominal variables used in previous research [22]. For this study, we considered appropriate to use 'strong' and 'moderate' correlation coefficients, as we believe it provides depth and richness. No 'very strong' correlations ($r_s > .8$) were found. 'Strong' and 'moderate' correlations were found among 46 variables (Table 1), which are considered as the most impactful ones among the previous set. For all cases, the correlations among the pairs of variables are significant as $p < .001$.

An undirected network of correlations was built (Fig 1.) and is composed of 46 nodes (variables) and 96 edges (correlations). The nodes size is ranked by total degree, i.e., the bigger the node (variable), the higher number of correlations with other variables (nodes). The edges thickness is ranked and labelled with the correlation coefficient ($r_s$). Nodes colours are partitioned according to the network modularity, and as observed, nine clusters were created. The pragmatic consistency of the clusters of variables was verified by the researchers and the clusters were then labelled according to their dominant logic. Table 1 depicts the labels of the variables for each cluster, the Mode of the student's answers, the CIPP element and the cluster label.

The cluster analysis reveals that the most relevant clusters belong to the contextual element of the CIPP model (K6, K5, K0 and K1), and that process elements are not totally linked to process; K3 'relationship with teachers/school' is, and K2 'relation-ship with peers' is not. The input element is totally isolated (K7) and liked to context the element regarding the (K0) 'family context'.

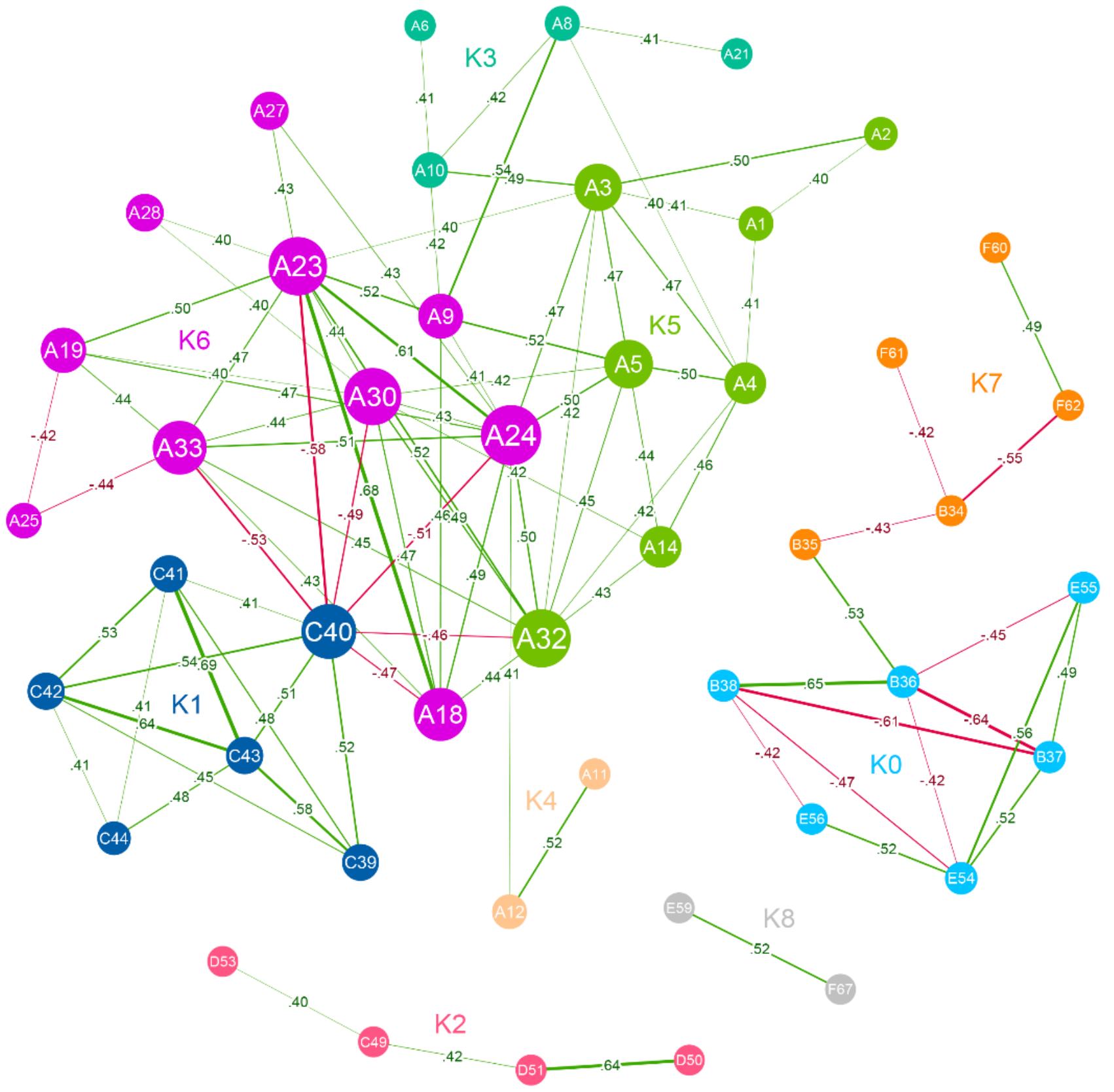

**Fig. 1.** The network of Spearman correlations (moderate and strong)

The cluster analysis also reveals that the sets of variables with higher correlation coefficient reside in the subdomains of the "Remote Learning Environment", "Pedagogy", "Interactions with teachers/school" and "Educational resources" (K6, K5, K3 and K4). These consists of the most impactful variables, according to students' responses.

**Table 1.** Depiction of variables' labels, clusters and mode

| K | Var | Mo | Variable label | CIPP / K label |
|---|---|---|---|---|
| 6 | A24 | 2 | In online classes, I learn as much as in face-to-face classes | Context / Remote learning environment |
| | A23 | 2 | I am motivated to participate in online classes | |
| | A30 | 2 | I am optimistic about my academic success this semester | |
| | A33 | 2 | After this phase, I will be more interested in taking classes/training online | |
| | A18 | 2 | Online classes help me to stay more positive during this phase | |
| | A9 | 3 | I've been feeling closer to my teachers | |
| | A19 | 3 | It's easy to work online with my classmates | |
| | A27 | 3 | Online classes are interactive, which allows me to participate | |
| | A28 | 4 | I have the necessary skills to use digital learning environments | |
| | A25 | 4 | Online classes are more tiring than face-to-face classes | |
| 5 | A32 | 3 | It has been possible to guarantee a minimum quality in my education | Context / Pedagogy |
| | A3 | 3 | My teachers have the necessary skills to manage online classes/work | |
| | A5 | 3 | The contact that my teachers keep with me is adjusted to my learning needs. | |
| | A4 | 3 | The classwork proposed by the teachers is appropriate | |
| | A14 | 3 | The online assessment mechanisms (tests, papers, etc.) are appropriate | |
| | A1 | 3 | My teachers created strategies to resume classroom work. | |
| | A2 | 3 | My teachers sent clear information about how online classes will work. | |
| 0 | B36 | 1 | At home, I need to share my workspace with other people, and this causes me limitations | Context / Family context |
| | E54 | 3 | My family environment facilitates my participation in online classes/work | |
| | B37 | 4 | At home, I have a proper workstation with the ideal conditions to participate in online classes/work | |
| | B38 | 2 | My work/study environment is disturbed by other people who live with me | |
| | E55 | 4 | In my family environment, my personal space and study/work schedules are respected | |
| | E56 | 4 | I feel cherished in my family environment | |
| 1 | C40 | 3 | I feel less motivated than usual | Context / Wellbeing |
| | C43 | 3 | I feel more nervous than usual | |
| | C41 | 3 | I feel more anxious than usual | |
| | C42 | 3 | I feel sadder than usual | |
| | C39 | 3 | I feel more tired than usual | |
| | C44 | 4 | I've been having more sleep disorders | |
| 7 | B34 | 4 | At home, I have all the necessary equipment for online classes/work | Input / Working conditions |
| | B35 | 1 | At home, I need to share the computer I work on, and this causes me limitations | |
| | F62 | 1 | I feel impeded from participating in online classes due to lack of adequate material/equipment | |
| | F61 | 1 | I needed to buy equipment/devices to participate in online classes/work | |
| | F60 | 3 | I am preoccupied with the worsening of my financial situation | |
| 2 | D51 | 3 | My wellbeing depends on keeping in contact with my friends/colleagues in school | Process / Social interactions with peers |
| | C49 | 1 | I have lost access to a safe space when the school closed | |
| | D50 | 4 | I miss my school friends/colleagues | |
| | D53 | 2 | The suspension of in-person classes has worsened my relationships with my friends/colleagues | |
| 3 | A10 | 3 | My teachers use software and digital educational resources suitable for online classes | Process / Social interactions with teachers/ school |
| | A8 | 3 | My teachers care about my personal wellbeing | |
| | A6 | 3 | My teachers have created alternative ways to contact me (chat rooms/groups, social networks, etc.) | |
| | A21 | 3 | The school board is committed to my welfare and academic success | |
| 4 | A12 | 2 | My teachers started using more creative and diverse resources/materials | Context / Educational resources |
| | A11 | 2 | I have gained access to more educational resources/materials | |
| 8 | F67 | 1 | The interruption of in-person classes has put my stay in higher education at risk, for financial reasons | Context / Risks |
| | E59 | 1 | The interruption of in-person school activities put my stay in higher education at risk, for family reasons | |

Within this group of clusters, the variables with a higher total degree are A24 “In online classes, I learn as much as in face-to-face classes”, and A23 “I am motivated to participate in online classes” which are positively and strongly correlated. The notion of learning as much in online classes tends to increase together with the motivation to participate in online classes. According to the previous results [22], the majority of the students answered “rarely” to both (c.f. column Mo in Table 1), thus, either students lacked the motivation to participate in online classes because they felt they did not learn as much or the other way around. Other possible motivational constrainers are found in variables correlated with A23, namely the rare belief that (A18) “Online classes help me to stay more positive during this phase”, the (A30) lack of optimism regarding their academic success in the confinement semester, and the students’ belief that after this phase they will rarely be (A33) “interested in taking classes/training online", which, in turn, is correlated with the almost consensual idea that (A25) "Online classes are more tiring than face-to-face classes".

The motivation to participate in online classes (A3) is also positively correlated with what we believe to be motivation enabling variables during the confinement period namely, the frequent notion that (A32) "It has been possible to guarantee a minimum quality in my education", from K5, the frequent notion of (A9) "feeling closer to my teachers", the frequent perception that (A27) "Online classes are interactive, which allows me to participate", the almost consensual assumption that students have (A28) "necessary skills to use digital learning environments". A3 is also positively correlated with the frequent belief that (A3) "My teachers have the necessary skills to manage online classes/work", from K5, and the frequent notion of (A19) being "easy to work online with my classmates", which is also negatively correlated with (A25) "Online classes are more tiring than face-to-face classes". Our pragmatic assumption is that the higher the easiness to work with peers online, the lower the feeling of tiredness in online learning.

The "Wellbeing" cluster (K1) is relatively isolated in the network but very close to the educational clusters, having a single entry point (C40) "I feel less motivated than usual", which is negatively correlated with six variables: A23, A33, A24, A30, A18, and A32. This means that unusual overall personal demotivation tends to increase or decrease when students feel or not motivated to participate in online classes (fostered by the previously mentioned enablers and constrainers), are more or less likely to take classes/training online after the confinement, are able or unable to learn as much in online classes than in face-to-face classes, are optimistic or not about their academic success, consider or not that online classes helped them to stay more positive during confinement, and consider or not that it was possible to guarantee the minimum quality of their education.

The intra-cluster analysis of K1 reveals that overall personal demotivation (C40) is positively correlated with (C42) sadness, (C39) tiredness, (C43) nervousness, and (C41) anxiety. There is a strong correlation triangle among nervousness, anxiety and sadness, while sadness and nervousness are also correlated with (C44) sleep disorders. Although tiredness is not correlated with sleep disorders, it is almost strongly correlated to nervousness.

The Social interactions cluster (K2) also appears relevant to the students' wellbeing with a strong correlation between (D51) "My wellbeing depends on keeping in contact with my friends/colleagues in school" (Mo=3) and (D50) "I miss my school friends/colleagues" (Mo=4). The remaining correlations suggest that if the relationship with friends/colleagues worsens, students are likely to feel that they lost access to a safe space.

In the Pedagogy cluster (K5), the most impactful variables consist of the guarantee of minimum quality in education (A32), the necessary skills that teachers have to manage online classes (A3), and the adjusted contact they keep with students. These variables appear positively correlated with the frequent notion of proposed classwork and assessment being adequate. The perception of the teachers' skills to manage online classes is also associated with the creation of strategies to resume classes during confinement and the clearness of regulatory information shared with students.

Regarding their family context (K0) and working conditions (K7), which we had previously grouped under the same dimension (Technological and working conditions), there is a clear separation between the organization of work/workload, privacy and disturbances, which students tend to relate to their family context (K0). In turn, the pure technological issues, such as owning, accessing, and sharing devices, are more closely linked to the students' financial situation and concerns (K7).

## 5 Discussion and conclusion

We make causation assumptions among correlations based on the students answers to each of the correlated variables since one of our objectives was to extract contextual knowledge about this specific sample in this specific period of analysis. For this, we used the modes of the answers for each variable combined with the correlation coefficient and the node total degree, to focus on the high impact variables on the network of correlations. The forwarded causation assumptions will most likely change for other datasets, though a certain level of abstraction in correlations may still be observed with other data for the same variables.

Results reveal the centrality of CIPP contextual elements and the centrality of motivation to participate in virtual classes during ERL as an important learning constrainer or enabler. Besides conditioning learning itself, motivation also shapes the students' perception of the role that online classes play in helping them to stay more positive during ERL. Students that were motivated and engaged showed a more positive attitude towards the ERL while those that were less motivated showed worst results. This is in line with the literature as authors [2] stress the importance of the context, in particular in unusual circumstances, which is the case. Additionally, it is also stressed that it is vital to learn how to develop empathy online as this contributes to involve all the actors in the learning process. Results also show that there is here some ground for future research, in particular concerning how to involve and engage students in online classes and virtual rooms as these are fundamental factors in the learning process.